\documentclass{article}
\usepackage{spconf,amsmath,graphicx}
\usepackage{algorithmicx}
\usepackage[noend]{algpseudocode}
\usepackage{booktabs}
\usepackage{multirow}
\usepackage{amsfonts}
\usepackage{color}
\usepackage{url}
\usepackage{enumitem}
\usepackage{makecell}

\title{Multi-task Siamese Neural Network for Improving Replay Attack Detection}
\name{Patrick von Platen$^{1,2}$, Fei Tao$^1$, Gokhan Tur$^1$}
\address{Uber AI$^1$ \\ RWTH Aachen University$^2$}

\begin{document}
%
\maketitle
\begin{abstract}
\vspace*{-0.2cm}
Automatic speaker verification systems are vulnerable to audio replay attacks which bypass security by replaying recordings of authorized speakers.
Replay attack detection (RA) detection systems built upon Residual Neural Networks (ResNet)s have yielded astonishing results on the public benchmark ASVspoof 2019 Physical Access challenge. With most teams using fine-tuned feature extraction 
pipelines and model architectures, the generalizability of such systems remains questionable though.
In this work, we analyse the effect of discriminative feature learning 
in a multi-task learning (MTL) setting can have on the generalizability and discriminability of RA detection systems.  
We use a popular ResNet architecture optimized by the cross-entropy criterion as our baseline and compare it to the same architecture optimized by MTL using Siamese Neural Networks (SNN).
It can be shown that SNN outperform the baseline by relative 26.8\% Equal Error Rate (EER). We further enhance the 
model's architecture and demonstrate that SNN with additional reconstruction 
loss yield another significant improvement of relative 13.8 \% EER. 
\end{abstract}

\vspace{-0.3cm}
\section{Introduction}
\label{sec:introduction}
\vspace{-0.3cm}

Automatic speaker verification (ASV) systems are nowadays increasingly used for various applications. However, ASV systems are vulnerable to \emph{audio spoofing attacks}, which attempt to gain unauthorized access by manipulating the audio input. One of the most popular and effective audio spoofing attacks are \emph{replay attacks} (RA)s. In an RA the attacker fools the ASV system by replaying a recording of an authorized speaker. Considering how effective and cheap RAs are, it is necessary to augment an ASV system with an RA detection system in practice.

The public benchmark ASVspoof initiative started with the ASVspoof 2015 challenge which dealt with text-to-speech and voice conversion spoofing attacks \cite{wu2015asvspoof}.
ASVspoof 2017 \cite{kinnunen2017asvspoof} was the first challenge concerned with RA detection and thus created a benchmark data set consisting of voice command recordings. ASVspoof 2019 \cite{todisco2019asvspoof}, then introduced a much larger 
corpus of longer and text-independent recordings for RA detection.

The performance of RA detection systems has been thought highly dependent on their input feature processing \cite{patil2018survey}.
Correspondingly, earlier work has largely dealt with handcrafted feature processing and it has been found that high frequency and phase information can be 
helpful for RA detection (\textit{e.g.} in \cite{nagarsheth2017replay, tom2018end}). 
Popular input features that emerged include linear frequency cepstral coefficients (LFCC) \cite{sahidullah2015comparison} and group delay (GD) grams \cite{hegde2006significance}. 
In recent years, input features derived from shorter handcrafted feature processing pipelines, such as the log power magnitude spectra (\emph{LOGSPEC}) \cite{lai2019assert}, attracted more interest. In contrast to LFCC, LOGSPEC preserves much more of the information present in the original raw signal and thus relies on deep neural netwokrs (DNN)s as powerful feature extractors \cite{lai2019assert, cai2019dku, gomez2019gated, lavrentyeva2019stc, zeinali2019detecting}. Overall, there is currently no conclusive consensus about the best input feature for RA detection. 

As the quality of recording and replaying devices is getting better, detecting the difference between genuine and spoofed audios is becoming more difficult. Thus, it becomes necessary to improve the discriminability and generalizability of RA detection systems. 
Besides common regularization techniques, like data augmentation and Dropout (\textit{cf.} with \cite{cai2019dku, zeinali2019detecting}), multiple teams have used discriminative loss functions and multi-task learning (MTL) \cite{caruana1997multitask} for better feature discrimination and generalization (\textit{cf.} with \cite{lai2019assert, lavrentyeva2019stc, jung2019replay}).

Siamese Neural Networks (SNN) \cite{bromley1994signature} have shown to significantly improve the discriminability and generalizability of models \cite{koch2015siamese}.
In this paper, we propose to use SNN in an MTL setting for RA detection. More generally, we investigate to what extent adding discriminative loss functions in a MTL setting can improve the performance of RA detection systems on the ASVspoof 2019 challenge Physical Access (PA) data. 
The analysis is conducted on multiple input features.
It is made sure that none of the systems rely on additional data and labels and that all of our settings follow the real-world application implementation. 
Our main contributions include: 1) Proposal of SNN in MTL setting for improved discriminability and generalizability of RA detection systems; 2) Extensive analysis of discriminative loss functions on multiple input features; 3) Enhancement of a popular architecture for RA detection with second-order statistics pooling; 4) Combination of reconstruction loss (ReL) with SNN in an MTL setting.

\vspace{-0.3cm}
\section{RELATED WORK}%
\label{sec:related_work}
\vspace{-0.3cm}

Convolutional neural networks (CNN)s and especially deep residual neural networks (ResNet)s \cite{he2016deep} have yielded the state-of-the-art performance on the ASVspoof 2019 PA data set \cite{lai2019assert, cai2019dku}. 
To deal with the much smaller data set than the one ResNet was originally designed for
\cite{lai2019assert, cai2019dku, jung2019replay} significantly reduce the size of their models by scaling down the number of kernels employed in each of the CNN layers. 
A key component in the architecture of their models is the projection of ResNet's three dimensional tensor output to a one dimensional vector 
for further binary classification.
In \cite{jung2019replay} a recurrent layer processes the tensor along the time dimension and outputs the last hidden state. A simpler and apparently more 
effective approach is to use a global average pooling (GAP) layer instead \cite{lai2019assert, cai2019dku}. Given the success of 
ResNet with GAP, we use this architecture as our baseline in this study. In other fields of research it has been shown that 
using second-order statistics in addition to first-order statistics yields
better feature embeddings for utterance level classification tasks, \textit{e.g.} in \cite{carreira2012semantic}. 
This led us to extent the GAP layer to additionally perform variance pooling.

In \cite{jung2019replay}, MTL has been applied for RA detection in the form of center loss (CL), which has been shown to greatly improve the discriminability of a model \cite{wen2016discriminative}. CL is comprised of the cross-entropy (CE) loss and the \emph{intra-class} variance loss of the feature embeddings weighted by a hyper parameter to control the 
intra-class compactness \cite{wen2016discriminative}.
SNN are known to significantly improve the discriminability and generalizability of a model \cite{koch2015siamese} and have found to be effective in similarity assessment in computer vision \cite{bell2015learning}.
By using a pair of input features during training, SNN simultaneously increase the \emph{inter-class} variance of the embedded input
features while decreasing the intra-class variance of the embedded input features.
Since CL can be seen as a special case of SNN \footnote{The centroid used in CL can be seen as one of the inputs in the input pair used for SNN.}, SNN are expected to better improve the discriminability of the model. This inspired us to propose SNN in a MTL setting for RA detection.

Another loss function, which is easily applicable in the MTL setting, is ReL. ReL is an unsupervised loss function and is 
usually employed in autoencoders to improve the network's ability to maintain the most distinctive information about the input features in compressed form. When added to a 
standard CE loss function, ReL can act as an effective regularizer by encouraging the network to learn robust feature embeddings \cite{reconstructLoss}. 

\vspace{-0.2cm}
\section{Proposed Approach}
\vspace{-0.2cm}
\subsection{Audio Preprocessing \& Feature Extraction}
\vspace{-0.2cm}

In a real-world application, the utterance input can be considered as a continuous buffer of audio input. We set the buffer size to $l_b$ seconds
to keep the audio processing step simple and easy to deploy.
Therefore, all utterances are cut or zero-padded to have 
a maximum length of $l_b$ seconds and \textbf{only} utterance-level input 
is considered.

The models are tested on the three input features: linear frequency filterbank features (\emph{LFBANK}), 
\emph{LOGSPEC} and \emph{GD grams}.
LFBANK correspond to the conventionally used LFCC features without the discrete cosine decorrelation step. 
We chose to leave out this decorrelation step because neural networks are known to act as excellent decorrelators.

We used modified GD grams as defined in Eqs. (28) and (29) in \cite{hegde2006significance} with $\alpha=0.4$ and $\gamma=0.9$ because the GD grams as formalized in \cite{tom2018end} and 
\cite{cai2019dku} did not yield any reasonable results in our experiments.
For all input features, the short time Fourier transform employed a window size of 50ms and a window shift of 15ms. 
LFBANK subsequently applies 80 filters (\textit{cf.} with \cite{chettri2019ensemble}) 
without any delta coefficients. The resulting input dimension for GD 
gram/LOGSPEC and LFBANK is $401 \times 566$ and $80 \times 566$, respectively.

Confirming the observations made in \cite{lai2019assert} and \cite{chettri2019ensemble}, the input features are \textbf{not} normalized, 
but simply scaled down to be in the range from $-1$ to $1$.

\vspace{-0.2cm}
\subsection{ResNet for Replay Attack Detection}
\vspace{-0.2cm}

Similar to \cite{cai2019dku}, the RA detection system is built upon a ''thin'' 34-layer ResNet, which is presented in detail in Table~\ref{tab:resnet_model}. The ResNet blocks (\textit{i.e.} \emph{Res1} - \emph{Res4}) employ the ''full pre-activation'' residual unit proposed in \cite{he2016identity}. 
Due to differences in the input dimensions between LOGSPEC/GD gram and LFBANK, slightly different stride kernels are used (\textit{cf.} with Table~\ref{tab:resnet_model}).
\begin{table}[ht]
\vspace{-0.4cm}
\caption{Architecture of ''thin'' 34-layer ResNet}
\begin{center}
\label{tab:resnet_model}
\begin{tabular}{@{} lcccc @{} } 
 \toprule
 Layer & Kernel & Filters & Input feature & Stride \\
 \midrule
  \multirow{2}{*}{Conv1} & \multirow{2}{*}{$[3 \times 3]$} & \multirow{2}{*}{16} &  GD/LOGSPEC & $[2 \times 2]$ \\
                        &&& LFBANK & $[2 \times 2]$ \\
 \midrule
   \multirow{2}{*}{Res1} & \multirow{2}{*}{$3 \times \begin{bmatrix} 3 \times 3 \\ 3 \times 3 \end{bmatrix}$} & \multirow{2}{*}{16} & GD/LOGSPEC & $[2 \times 2]$ \\
                    &&& LFBANK & $[1 \times 1]$ \\
 \midrule
 \multirow{2}{*}{Res2} & \multirow{2}{*}{$4 \times \begin{bmatrix} 3 \times 3 \\ 3 \times 3 \end{bmatrix}$} & \multirow{2}{*}{32} & GD/LOGSPEC & $[2 \times 2]$ \\
                    &&& LFBANK & $[1 \times 2]$ \\
 \midrule
 \multirow{2}{*}{Res3} & \multirow{2}{*}{$6 \times \begin{bmatrix} 3 \times 3 \\ 3 \times 3 \end{bmatrix}$} & \multirow{2}{*}{64} & GD/LOGSPEC & $[1 \times 1]$ \\
                    &&& LFBANK & $[2 \times 2]$ \\
 \midrule
 \multirow{2}{*}{Res4} & \multirow{2}{*}{$3 \times \begin{bmatrix} 3 \times 3 \\ 3 \times 3 \end{bmatrix}$} & \multirow{2}{*}{128} & GD/LOGSPEC & $[1 \times 1]$  \\
                    &&& LFBANK & $[2 \times 2]$ \\
\bottomrule
\vspace{-0.6cm}
\end{tabular}
\end{center}
\end{table}
The ResNet network is followed by a GAP layer as explained in Eq. (1) in \cite{cai2019dku}. Extending GAP to the retrieval of second-order statistics, we define a global average and variance pooling (GAVP) layer that extracts both the mean and variance from all feature maps of ResNet's last CNN layer. 
To keep the number of parameters constant, the pooling layer is followed by a $128 \times 64$ dense layer if GAP is employed and a $256 \times 32$ dense layer if GAVP is employed. The final dense layer (called ''Out'' in Fig.~\ref{fig:snn_all}) following the GAP or GAVP layer has a \textbf{single output 
neuron} with Sigmoid activation function yielding the probability of the input feature to be classified as being \emph{spoofed}.
All layers except the pooling and final layer make use of the Rectified Linear Unit (ReLU).
The model has about 1.34 million trainable parameters. 

\vspace{-0.2cm}
\subsection{Multi-Task Learning with Siamese Neural Networks}
\label{sub:snn}
\vspace{-0.2cm}

\begin{figure}[h]
    \centering
    \includegraphics[scale=0.3]{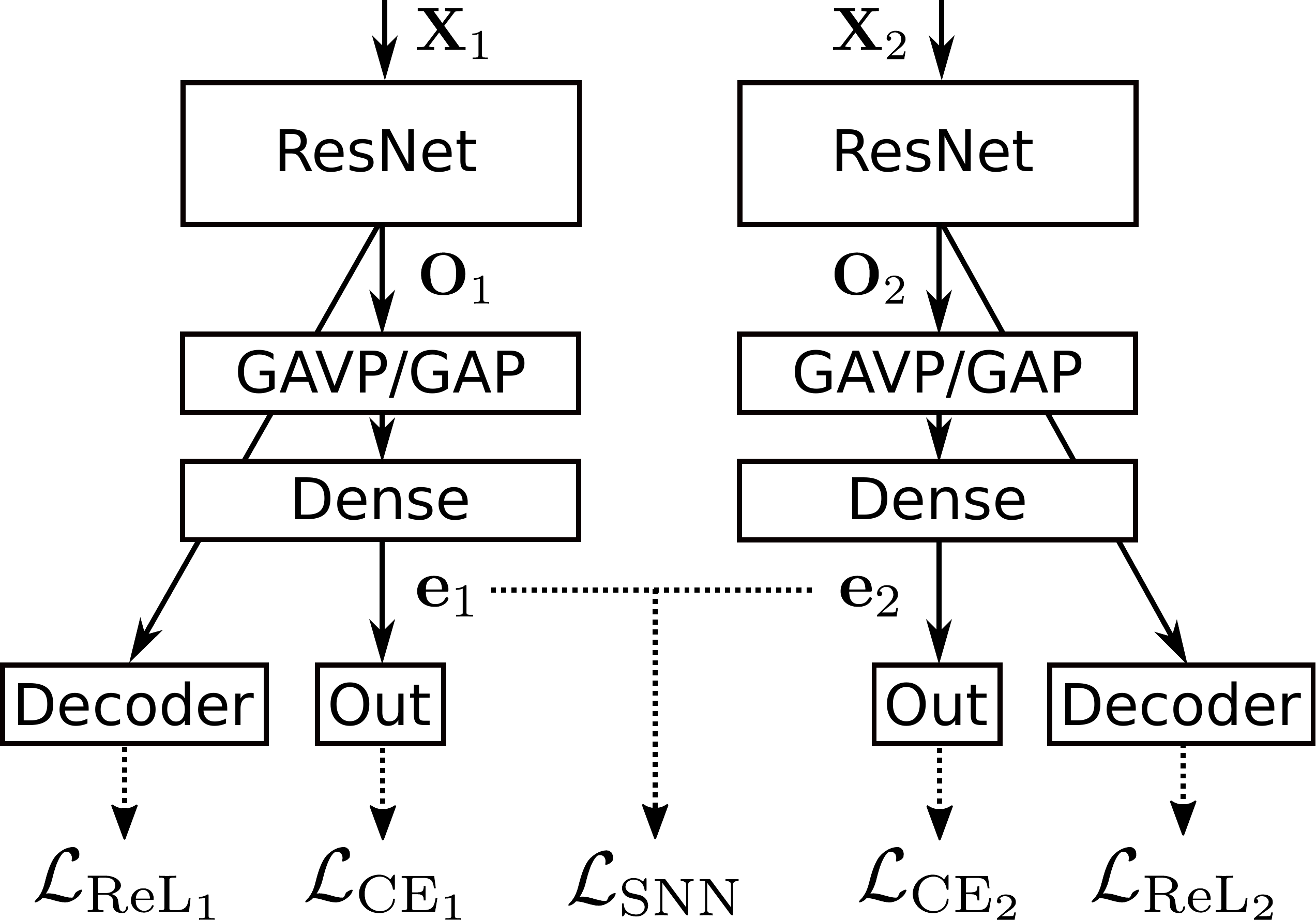}
\vspace{-0.3cm}
    \caption{A sketch map of SNN for MTL showing the loss functions $L_{\text{CE}_1}, L_{\text{CE}_2}, L_{\text{SNN}}$ and $L_{\text{ReL}_1}, L_{\text{ReL}_2}$ for RA detection.}
    \label{fig:snn_all}
\end{figure}

SNN are made of two sub-networks which share the same set of trainable parameters so that a pair of input features $\mathbf{X}_1,\mathbf{X}_2$ is used as an input during training. 
Besides computing the conventional CE loss for each sub-network individually (\textit{i.e.} $\mathcal{L}_{\text{CE}_1}$, $\mathcal{L}_{\text{CE}_2}$), a distance loss $\mathcal{L}_{\text{SNN}}$ between the feature embedding (\textit{i.e.} $\mathbf{e}_{1}, \mathbf{e}_{2}$) of each sub-network is calculated (\textit{cf.} with Fig.~\ref{fig:snn_all}).
A common choice for $\mathcal{L}_{\text{SNN}}$ is the hinge loss (\textit{cf.} with \cite{rosasco2004loss}): 
\begin{equation}\label{eq:siamese_loss}
    \mathcal{L}_{\text{SNN}} = \underset{(\mathbf{X}_1,
    y_1),(\mathbf{X}_2, y_2) \sim \mathcal{D} \times \mathcal{D}}
    {\mathbb{E}}
    [\max(0, m - l_dd(\mathbf{e}_1,\mathbf{e}_2))],
\end{equation}
wheres $m$ represents the margin, $l_d$ equals $1$ if the input feature labels $y_1, y_2$ are equal or else $-1$ and $d(\mathbf{e}_1,\mathbf{e}_2)$ is a distance metric of choice for which we empirically found the 
cosine similarity to work best. During training, SNN then aims at minimizing the sum of $\mathcal{L}_{\text{CE}_1}$, $\mathcal{L}_{\text{CE}_2}$ and $\mathcal{L}_{\text{SNN}}$, whereas each loss contributes with equal weight.

Optionally, two ReLs ($\mathcal{L}_{\text{ReL}_1}, \mathcal{L}_{\text{ReL}_2}$) - one for each sub-network - can be added 
to the overall loss. In this case a shared decoder (with a negligible amount of parameters) is used to 
reconstruct the pair of input features $\mathbf{X}_1, \mathbf{X}_2$ from the outputs $\mathbf{O}_1, \mathbf{O}_2$ of the last convolutional layer:
\begin{equation} \label{eq:reconstruct_loss}
    \mathcal{L}_{\text{ReL}_i} = |\mathbf{X}_i - \text{Decoder}(\mathbf{O}_i)|_{F}^2 \text{ ,} \forall i \in \{1,2\},
\end{equation}
with $|\cdot|_{F}$ being the Frobenius norm. The decoder consists of three consecutive Deconvolution layers each of which upsamples the input using the stride kernel $[ 2 \times 2 ]$ and which employ $32,16,8$ kernels of size $[3 \times 3]$ respectively. 
The outputs of $\text{Decoder}(\mathbf{O}_1)$ and
$\text{Decoder}(\mathbf{O}_2)$ are ''mean-pooled'' over their output
feature maps and finally zero-padded to have exactly the same dimension
as their respective input feature matrices. The complete architecture 
of SNN is illustracted in Fig.~\ref{fig:snn_all}.

As can be noted from Eq. \eqref{eq:siamese_loss}, the space of possible training samples for SNN includes
all pair-wise combinations of $\mathcal{D}$ with itself, which is prohibitively large. A simple remedy 
taken in this study is to control the dataset's size by a hyper parameter \emph{numSamples}. Before every epoch, a dataset $\mathcal{D}_{SNN}$ is created by the following simple, but effective sampling procedure:
\begin{algorithmic}
\Function{CreateSNNDataSet}{$numSamples$}
\State $\mathcal{D}_{pos}, \mathcal{D}_{neg} \gets \textsc{splitByLabelAndShuffle}(\mathcal{D})$
\State ${c}_{pos}, {c}_{neg} \gets 0$
\For{$i$ \textbf{in} $1$ \textbf{to} $numSamples$}
\For{$j$ \textbf{in} $1$ \textbf{to} $2$}
    \State $\mathcal{D}_{r}, {c}_{r} \gets \textsc{rand}((\mathcal{D}_{pos}, {c}_{pos}), (\mathcal{D}_{neg}, {c}_{neg}))$
    \State $\mathbf{X}_{j},y_{j} \gets \mathcal{D}_{r}[c_{r}], \mathcal{D}_{r} == \mathcal{D}_{pos}$
    \State ${c}_{r} \gets ({c}_{r} + 1) \bmod \textsc{len}(\mathcal{D}_{r})$
    \EndFor
    \State $\mathcal{D}_{SNN}[i] \gets ((\mathbf{X}_1, y_1), (\mathbf{X}_2, y_2))$
\EndFor
\State \Return $\mathcal{D}_{SNN}$
\EndFunction
\end{algorithmic}
First, no data sample in $\mathcal{D}_{pos}$ (or $\mathcal{D}_{neg}$, respectively) is used twice \emph{before} every other data sample has been sampled at least once, which ensures almost certainly that all data samples are used per epoch by setting \emph{numSamples} accordingly. 
Second, it is ensured that the space of possible sample pairs $\mathcal{D} \times \mathcal{D}$ is vastly explored by shuffling the order of $\mathcal{D}_{pos}$ and $\mathcal{D}_{neg}$ 
before every epoch. Third, by choosing to sample from $\mathcal{D}_{pos}$ or $\mathcal{D}_{neg}$ with even 
probability, the smaller of the two data sets is upsampled so that $\mathcal{D}_{SNN}$ is balanced. 

\vspace{-1em}

\vspace{-0.2cm}
\section{Experimental Setup and Results}
\vspace{-0.3cm}

In all experiments, the models were evaluated on the PA subset of 
the ASVspoof 2019 corpus \cite{todisco2019asvspoof}. 
PA consists of 48600 \emph{spoofed} and 5400 \emph{genuine} utterances
in the training (\emph{train}) data set, 24300 \emph{spoofed} and 5400
\emph{genuine} utterances in the development (\emph{dev}) data set and
116640 \emph{spoofed} and 18090 \emph{genuine} utterances in 
the evaluation (\emph{eval}) data set.
The models were optimized by Adam with $\beta_1=0.9$, $\beta_2=0.999$,
learning rate $3.95 \times 10^{-4}$ and weight decay, which was tuned for each experiment separately. Training was stopped if the equal error rate (EER) on the dev data set did not improve over 15 consecutive epochs. The models were implemented with the Keras framework \cite{chollet2015keras}.

First, we analysed the effect of audio input length on the performance of a simplified ResNet model using LFBANK on the eval set. We noticed that increasing the input length $l_b$ from 5.0s to 6.5s and eventually to 8.5s improved the EER from 9.31 \% to 6.75 \% to finally 6.22 \%. In this experiment, we simply cut or padded the end of the audio to the specific length. Based on existing literature (\textit{e.g.} \cite{chettri2019ensemble}), it can be explained that the beginning and tailing silence cues can lead to better performance. Considering these findings and our practical application, we decided to use 8.5s input length and to do cutting and padding at the end of the audio from now on (so that we do not rely on voice activity detection in practical applications).

We then analyse the proposed model architecture with GAP \cite{cai2019dku}. As one baseline, the model was trained using simple CE loss ($\mathcal{L}_{\text{CE}}$) \cite{jung2019replay}, which is abbreviated as $M_{\text{CE}}^{\text{GAP}}$. 
As another baseline, the model was trained using CL loss ($\mathcal{L}_{\text{CE}} + \gamma \mathcal{L}_{\text{CL}}$), which is abbreviated as $M_{\text{CL}}^{\text{GAP}}$. For $M_{\text{CL}}^{\text{GAP}}$, we found that $\gamma=0.001$
yields the best results. The baselines are compared to SNN as described in Section~\ref{sub:snn}, which we abbreviate as $M_{\text{SNN}}^{\text{GAP}}$. 
For $M_{\text{SNN}}^{\text{GAP}}$ the \emph{numSamples} was set to $1 \times 10^6$ and the margin $m$ was set to $0.5$. In all training setups, we used a batch size of $32$.
$M_{CE}^{GAP}, M_{CL}^{GAP}$ and $M_{\text{SNN}}^{\text{GAP}}$ were all evaluated using LFBANK, LOGSPEC and GD gram as input. In a final step, the systems were systematically fused by means of logistic regression with the Bosaris toolkit \cite{brummer2013bosaris} using the \emph{dev} data set for calibration. 

Due to the data imbalance of 9 to 1 in the training set, we adopted the weighted CE loss for $M_{\text{CE}}^{\text{GAP}}$ and  $M_{\text{CL}}^{\text{GAP}}$ with the CE weight for \emph{spoofed} utterance input  
set to $\frac{1}{9}$. To improve training stability, the bias of the output neuron was initialized to $\log(9)$ (\textit{cf.} with \cite{lin2017focal}) if weighted CE was used.
The results can be seen in Table~\ref{tab:experiment_1}.

\vspace{-1em}
\begin{table}[ht]
\caption{Comparison of $M_{\text{CE}}^{\text{GAP}}, M_{\text{CL}}^{\text{GAP}}, M_{\text{SNN}}^{\text{GAP}}$ for different input features. Results are reported in \% EER.}
\begin{center}
\label{tab:experiment_1}
\begin{tabular}{@{} lcccc @{} } 
 \toprule
 Model & Loss & Input Feature & Dev & Eval \\
 \midrule
 \multirow{4}{*}{$M_{\text{CE}}^{\text{GAP}}$} &
 \multirow{4}{*}{$\mathcal{L}_{\text{CE}}$} & 
       LFBANK & 3.70 & 5.17 \\
    && GD Gram & 6.20 & 8.63 \\
    && LOGSPEC & 1.98 & 2.79 \\
    && Fused & - & 2.22 \\
 \midrule
  \multirow{4}{*}{$M_{\text{CL}}^{\text{GAP}}$} & \multirow{4}{*}{\makecell{$\mathcal{L}_{\text{CE}} +$ \\ $ 0.001\mathcal{L}_{\text{CL}}$}} & 
                           LFBANK & 2.76 & 4.06 \\
                        && GD Gram & 4.44 & 7.13 \\
                        && LOGSPEC & 1.37 & 2.33 \\
                        && Fused & - & 1.70 \\
  \midrule
  \multirow{4}{*}{$M_{\text{SNN}}^{\text{GAP}}$} & \multirow{4}{*}{\makecell{$\mathcal{L}_{\text{CE}_1} + $ \\ $ \mathcal{L}_{\text{CE}_2} + \mathcal{L}_{\text{SNN}}$}} &
                           LFBANK & 2.73 & 3.66 \\
                        && GD Gram & 3.53 & 5.89 \\
                        && LOGSPEC & \textbf{1.15} & \textbf{2.25} \\
                        && Fused & - & 1.52 \\
\bottomrule
\end{tabular}
\end{center}
\end{table}
\vspace{-1em}

It can be seen that both MTL models $M_{\text{CL}}^{\text{GAP}}$ and $M_{\text{SNN}}^{\text{GAP}}$ 
outperform the single task learning model $M^{\text{GAP}}_{\text{CE}}$ by relative 23.4 \% EER and 31.5 \% EER averaged over all input features. $M_{\text{SNN}}^{\text{GAP}}$ further 
outperforms $M_{\text{CL}}^{\text{GAP}}$ by a relative margin of 10.6 \% EER. We could 
observe that during training, the MTL setups $M_{\text{CL}}^{\text{GAP}}$ and $M_{\text{SNN}}^{\text{GAP}}$ converged faster and also seemed to generalize better as the EER on the dev data set decreased much smoother during training. 

In the second experiment, we took the best performing model $M^{\text{GAP}}_{\text{SNN}}$ for LOGSPEC input as our new baseline.
First, we analysed the effect of extracting second-order statistics 
in addition to first-order statistics from of the CNN feature maps by 
replacing the GAP layer with a GAVP layer. This setup is abbreviated as 
$M^{\text{GAVP}}_{\text{SNN}}$. Second, we extended SNN with two additional 
reconstruction loss functions $\mathcal{L}_{\text{ReL}_1}, \mathcal{L}_{\text{ReL}_2}$ according to Eq.~\eqref{eq:reconstruct_loss} for 
both GAP ($M^{\text{GAP}}_{\text{SNN},\text{ReL}}$) and GAVP ($M^{\text{GAVP}}_{\text{SNN},\text{ReL}}$). Empirically, it was found
that $\mathcal{L}_{\text{ReL}_1}$ and $\mathcal{L}_{\text{ReL}_2}$ is
much smaller than $\mathcal{L}_{\text{CE}_1},
\mathcal{L}_{\text{CE}_2}$ and $\mathcal{L}_{\text{SNN}}$, so that the
loss is scaled by a weighting factor of $50$. 
Because, we experienced RAM memory overflow issues 
with $M^{\text{GAP}}_{\text{SNN},\text{ReL}}$ and $M^{\text{GAVP}}_{\text{SNN},\text{ReL}}$, the batch size used in training was reduced to $16$ and \emph{numSamples} set to $5 \times 10^{5}$ 
to have the same number of steps per epoch as before 
\footnote{More details can be found at \url{https://www.comet.ml/patrickvonplaten/anti-spoof}.}.
The results are shown in Table~\ref{tab:experiment_2}.

\vspace{-1em}
\begin{table}[ht]
\caption{Comparison of $M_{\text{SNN}}^{\text{GAVP}}, M_{\text{SNN},\text{ReL}}^{\text{GAP}}$ and $M_{\text{SNN},\text{ReL}}^{\text{GAVP}}$ for LOGSPEC input feature. Results are reported in \% EER.}
\begin{center}
\label{tab:experiment_2}
\begin{tabular}{@{} lccc @{} } 
 \toprule
 Model & Loss & Dev & Eval \\
 \midrule
  $M_{\text{SNN}}^{\text{GAVP}}$ & \makecell{$\mathcal{L}_{\text{CE}_1} + $ \\ $ \mathcal{L}_{\text{CE}_2} + \mathcal{L}_{\text{SNN}}$} & 0.83 & 2.01 \\
  \midrule
    $M_{\text{SNN},\text{ReL}}^{\text{GAP}}$ & \makecell{$\mathcal{L}_{\text{CE}_1} + \mathcal{L}_{\text{CE}_2} + \mathcal{L}_{\text{SNN}} $ \\ $ + 50\mathcal{L}_{\text{ReL}_1} + 50\mathcal{L}_{\text{ReL}_2}$} & \textbf{0.66} & 2.08 \\
  \midrule
    $M_{\text{SNN},\text{ReL}}^{\text{GAVP}}$ & \makecell{$\mathcal{L}_{\text{CE}_1} + \mathcal{L}_{\text{CE}_2} + \mathcal{L}_{\text{SNN}} $ \\ $ + 50\mathcal{L}_{\text{ReL}_1} + 50\mathcal{L}_{\text{ReL}_2}$} & 0.76 & \textbf{1.94} \\
\bottomrule
\end{tabular}
\end{center}
\end{table}
\vspace{-1em}

It can be seen that both using the GAVP layer 
and adding ReL gives a significant performance boost compared to $M_{\text{SNN}}^{\text{GAP}}$. Consequently the best 
single system performance of $1.94$ \% EER on the \emph{eval} data set is achieved 
by $M_{\text{SNN},\text{ReL}}^{\text{GAVP}}$ which outperforms $M_{\text{M}_{CE}^{GAP}}$ by relative 
30.5 \% EER while having the same number of parameters. 

\vspace{-0.3cm}
\section{Conclusion}
\vspace{-0.3cm}
\label{sec:conclusion}

We have thoroughly analysed the discriminate feature learning in 
an MTL setting for RA detection and found that SNN significantly 
outperforms the baseline on multiple input features. 
We explain this improvement by the following. 
First, SNN greatly improve the discriminability of the model by \emph{explicitly} increasing the inter-class variance of the model. Second, because SNN sample from a very large pool of possible  
sample pairs - each giving a different gradient signal - the model regularizes much better during training. 
We then further improve upon SNN by adding ReL and replacing GAP with GAVP. This leads to a single system EER of $1.94$\% and can be
justified by better regularization induced by ReL and more 
discriminative feature embeddings thanks to the extraction of first- and second-order statistics.

\bibliographystyle{IEEEbib}
\section{References}
{\footnotesize
\bibliography{refs}}

\end{document}